\documentstyle[prc,aps]{revtex}

\newcommand{\newc}{\newcommand}
\newc{\be}{\begin{equation}}
\newc{\ee}{\end{equation}}
\newc{\bea}{\begin{eqnarray}}
\newc{\eea}{\end{eqnarray}}
\begin{document}
\title{Neutrinoless Double Beta Decay of $^{134}Xe$ }
\author{ 
Fedor \v Simkovic$^{1,2)}$,
Pavol Domin$^{2,3)}$
and Amand Faessler$^{1)}$
}
\address{
$^1$  Institute of Theoretical Physics, University of Tuebingen,
D--72076 Tuebingen, Germany\\
$^2$Department of Nuclear Physics,  
Comenius University, 
SK--842 15 Bratislava, Slovakia\\
$^3$ Laboratory of Nuclear Problems, Joint Institute for Nuclear 
Research, 141980 Dubna, Russia 
}
\date{\today}
\maketitle
\draft
\begin{abstract}
{In view of recent great progress achieved in the experimental study of 
the neutrinoless double beta decay ($0\nu\beta\beta$-decay) of
$^{134}Xe$ we discuss theoretical aspects of this process. 
The light and heavy Majorana neutrino exchange
as well as the trilinear R-parity breaking
contributions to $0\nu\beta\beta$-decay are considered.
We show that the sensitivity of the studied process to the 
signal of lepton number violation is only by factor 2-3
weaker in comparison with the $0\nu\beta\beta$-decay of
$^{136}Xe$. The current limits on effective neutrino mass
(light and heavy) and trilinear R-parity violating
parameter ${\acute{\lambda}}_{111}$ deduced
from lower limits on the $0\nu\beta\beta$-decay 
half-lifes of various nuclei are reviewed
and perspectives of the experimental verification of  
recently announced evidence of the $0\nu\beta\beta$-decay 
of $^{76}Ge$ are discussed.

}
\end{abstract}

The neutrinoless double beta decay ($0\nu\beta\beta$-decay)
does exist if the neutrino is a Majorana particle with non-zero mass 
\cite{sch82}. Many extensions of the Standard model (SM) breaks lepton 
number (LN), i.e., generate Majorana neutrino masses,  
and offer a plethora of  $0\nu\beta\beta$-decay mechanisms 
mediated by a variety of virtual particles, in particular
by exchange of light and heavy neutrinos, supersymmteric (SUSY) 
particles, leptoquarks etc (for a review see Refs. 
\cite{fae98,vog02,ver02,klapd,eji00}). In 
connection with the observation of oscillations of solar and 
atmospheric neutrinos most attention
is concentrated to the neutrino mass mechanism of the
$0\nu\beta\beta$-decay due to possibility to predict 
the mass spectrum of the neutrinos \cite{bil99}. However, the
observed absence of the $0\nu\beta\beta$-decay 
allows to put stringent constraints also on many other parameters
of modern  extensions of the SM like Grand
unified models and their SUSY versions. 

There are more than 50 even-even nuclei for which both $\beta^-$  and
$\beta^+$ decays  are forbidden energetically, 
i.e., can be used for experimental investigation of the 
$0\nu\beta\beta$-decay \cite{data}. Between them one can find twin isotopes
$^{128,130}Te$ and $^{134,136}Xe$, which merely differ by two neutrons
and are of comparable abundance in the nature. The first pair of 
isotopes has been discussed often in the literature because 
of geochemical measurements and the Pontecorvo argument that 
the ratio of  corresponding half-lifes is essentially independent
of nuclear physics. 
A small difference of the corresponding $0\nu\beta\beta$-decay 
nuclear matrix elements of the order of 10-20\%  has been found 
in nuclear structure studies \cite{si99,woda}. If the 
situation would be similar also in the case of xenon isotopes, the 
importance of the experimental study of the $0\nu\beta\beta$-decay 
for $^{134}Xe$ would be minor due to a significantly smaller Q-value 
in comparison with that for $^{136}Xe$, as it implies weaker
bounds on the LN violating parameters. A detail theoretical
study of this $0\nu\beta\beta$-decay channel is missing. There is 
only one prediction for
the light neutrino mass mechanism of this process \cite{muto}.

Recently, a new limit on the half-life of the $0\nu\beta\beta$-decay 
of $^{134}Xe$ to the ground state with 
$T^{0\nu-exp}_{1/2} \ge 5.8\times 10^{25}$ has been obtained 
at Gran Sasso by means of the 6.5 kg liquid xenon setup of the
DAMA experiment \cite{xe134,xe136}, which is 
more stringent  in comparison with the
previous one by about three orders of magnitude \cite{barab}.
The aim of this contribution is to analyze the light and heavy Majorana
exchange and the trilinear R-parity breaking ($R_p \hspace{-1em}/\;\:$)
mechanisms for this rare process. A realistic 
calculation of nuclear matrix elements will be performed within the 
renormalized Quasiparticle Random Phase Approximation (RQRPA) \cite{simn96}. 
A comparison with the $0\nu\beta\beta$-decay of $^{136}Xe$ will
be presented.  The sensitivity of different 
$0\nu\beta\beta$-decay experiments looking for signal of LN
violation will be compared. In addition, a possible 
experimental verification of the evidence for the 
$0\nu\beta\beta$-decay of $^{76}Ge$ will be discussed. 

The half-life of the $0\nu\beta\beta$-decay associated with
light and heavy Majorana neutrino mass mechanism is given as
\begin{equation}
[T_{1/2}^{0\nu}]^{-1} = G_{01} 
|\frac{<m_\nu >}{m_e} M^{light}_{<m_\nu >} + 
\eta_{_{N}} M^{heavy}_{\eta_{_N}}|^2.
\label{eq:1}
\end{equation}
The effective light and inverse heavy Majorana masses are
\begin{eqnarray}
<m_\nu > ~ &=& ~ \sum^{light}_k~ (U^L_{ek})^2 ~ \xi_k ~ m_k, 
\nonumber \\
\eta_{_N} ~ &=& ~ \sum^{heavy}_k~ (U^L_{ek})^2 ~ 
{\hat \xi}_k ~ \frac{m_p}{M_k}.
\label{eq:2}   
\end{eqnarray}
Here,  $m_p$ ($m_e$)  is the proton (electron)  mass, 
$U^L_{ek}$ mixing matrix elements and $\xi_k, \hat{\xi_k}$
phases associated with the charge conjugation of the Majorana
neutrino field. $G_{01}$ denotes the phase space factor given in
\cite{doi83,pan96}.  The nuclear matrix elements $M_{<m_{ee}>}$ 
and $M_{\eta_{_N}}$ can be written as
a sum of Fermi, Gamow-Teller and tensor contributions \cite{si99}
\begin{equation}
M_{\cal{I}} = -\frac{M^{\cal{I}}_F}{g^2_A} + M^{\cal{I}}_{GT} 
+ M^{\cal{I}}_T
\label{eq:3}   
\end{equation}
with ${\cal{I}} = {<m_{ee}>}, \eta_{_{N}}$ and  $g_A = 1.25$. 
It is worthwhile to notice that the above matrix elements include 
contribution from induced nucleon currents. Recently, it
has been found that it is significant  and leads to a modification
of Gamow-Teller and to new tensor contributions \cite{si99}.

The $0\nu\beta\beta$-decay half-life associated 
with $R_p \hspace{-1em}/\;\:$ SUSY
mechanism mediated by exchange of gluinos \cite{fae98,woda,awf99}
takes the form
\begin{equation}
\big[ T_{1/2}(0^+ \rightarrow 0^+) \big]^{-1} = 
G_{01}~ 
\left[
\frac{\pi \alpha_s}{6}
\frac{\lambda^{'2}_{111}}{G_F^2 m_{\tilde d_R}^4}
\frac{m_p}{m_{\tilde g}}\left(
1 + \left(\frac{m_{\tilde d_R}}{m_{\tilde u_L}}\right)^2\right)^2
\right]^2
|{\cal{M}}_{{\acute{\lambda}}_{111}}|^2.
\label{eq:4}   
\end{equation}
Here, $G_F$ is the Fermi constant and $\alpha_s = g^2_3/(4\pi )$ denotes 
 $\rm SU(3)_c$ gauge coupling constant.
$m_{{\tilde u}_L}$, $m_{{\tilde d}_R}$ and  $m_{\tilde g}$
are masses of the u-squark, d-squark and  gluino, respectively.
The nuclear matrix element ${\cal{M}}_{{\acute{\lambda}}_{111}}$
can be written as sum of  one- and two- pion exchange contributions:
\begin{equation}
{\cal M}_{{\acute{\lambda}}_{111}} = c_A \Big[
 \frac{4}{3}\alpha^{1\pi}\left(M_{GT}^{1\pi} + M_{T}^{1\pi} \right)
      +
      \alpha^{2\pi}\left(M_{GT}^{2\pi} + M_{T}^{2\pi} \right)\Big]\,
\label{eq:5}
\end{equation}
with $c_{_{A}} = m^2_{_{A}}/(m_p m_e)$ ($m_A = 850$ MeV). 
The structure coefficients $\alpha^{1\pi, 2\pi}$ for the one-pion 
and two-pion exchange contributions are given in \cite{awf99}.

The nuclear matrix elements for light and heavy Majorana neutrino
exchange and $R_p \hspace{-1em}/\;\:$ gluino exchange in
the $0\nu\beta\beta$-decay of $^{134}Xe$ are given
in Table \ref{table.1}. They have been obtained 
within the proton-neutron RQRPA \cite{simn96} for a model
space comprising the full $2-5\hbar\omega$ major shells 
like in previous studies of the $0\nu\beta\beta$-decay of $^{136}Xe$.
The details of the nuclear structure approach and the evaluation of
the matrix elements can be found in Refs. \cite{si99,woda}.  
By glancing to Table \ref{table.1} we see that values of 
matrix elements for A=134 are considerably larger (by factor 2-3) 
in comparison with those for A=136. Thus a pair of xenon isotopes
offers a different scenario from  the pair of 
tellurium nuclei. As it was discussed in previous publications
\cite{si99,exfs} the reason of the suppression  of
the $0\nu\beta\beta$-decay transition $^{136}Xe \rightarrow {^{136}}Ba$
is that the parent nucleus is a closed shell nucleus for neutrons (N=82).
The Pauli blocking effect is strongly suppressing some of
the proton-particle neutron-hole Gamow-Teller configurations, in particular
$(0g_{7/2,9/2},0g_{7/2,9/2})$ and $(1d_{3/2,5/2},0g_{7/2,9/2})$.

In order to compare the 
 sensitivities to the signal of LN violation of the considered 
$0\nu\beta\beta$-decays for xenon isotopes the 
corresponding kinematical factors $G_{01}$  
have to be known. Their values are determined by the mass difference
between the mother and daughter nuclei. They are $3.479~MeV$ and
$1.841~MeV$ for A = 136 and 134, respectively. 
One finds (assuming $r_0~=~1.1~fm$): $G_{01}~=~5.91\times 10^{-14}~(A=136)$, 
$2.30\times 10^{-15}~(A=134)$. The sensitivity parameters  with respect to 
different LN violating scenarios have been introduced in
previous works \cite{si99,awf99,erice} and are given by 
\begin{eqnarray}
\zeta_{<m_\nu >} (Y) & = &
10^{7}~ |{\cal M}^{}_{<m_\nu >}|~ 
\sqrt{{G_{01}}~ {year}},\nonumber \\
\zeta_{\eta_{_N}} (Y)  &=&  
10^{6}~ |{\cal M}^{}_{\eta_{_N}}|~ \sqrt{{G_{01}}~{year}},
\nonumber \\
\zeta_{\lambda_{111}'} (Y) &=& 
10^{5}~ |{\cal M}^{}_{\lambda_{111}'}|~ 
\sqrt{{G_{01}}~{year}}.
\label{eq:6}   
\end{eqnarray} 
For nuclei of experimental interest their values are presented in
Table \ref{table.2}. We see that the sensitivity of 
$0\nu\beta\beta$-decays of $^{134}Xe$ to the effective light
Majorana mass $<m_\nu >$ is only by factor 2 smaller in comparison
with that for  $^{136}Xe$. An interesting moment is that in the case
of heavy Majorana neutrino and $R_p \hspace{-1em}/\;\:$ SUSY
modes the corresponding ratio of the sensitivity parameters is larger
by about a factor 3. So, if in the xenon $0\nu\beta\beta$-decay experiment
half-lifes for both isotopes will be measured, one can draw conclusions
about the dominance of one of the mechanisms
of this process. A necessary condition for it is 
that the nuclear matrix elements governing this process
are calculated correctly. 

Knowing the values of sensitivity parameters it is straightforward
to deduce upper bounds on the effective LN violating 
parameters from the experimental lower limits on the 
$0\nu\beta\beta$-decay half-life $T^{0\nu -exp}_{1/2}$
for a given nucleus:
\begin{eqnarray}
\label{eq:7}
\frac{<m_\nu >}{m_e} &\leq& \frac{10^{-5}}{\zeta_{<m_\nu >}}
\sqrt{\frac{10^{24}~years}{T^{0\nu -exp}_{1/2}}}, ~~~~~~~
\eta_{_N} \leq \frac{10^{-6}}{\zeta_{\eta_{_N}}}
\sqrt{\frac{10^{24}~years}{T^{0\nu -exp}_{1/2}}}, \nonumber \\
(\lambda_{111}' )^2 &\leq&
\kappa^2~\left( \frac{m_{\tilde{q}}}{100~GeV}\right)^4~
\left(\frac{m_{\tilde{g}}}{100~GeV}\right)
\frac{10^{-7}}{\zeta_{\lambda_{111}'}}
\sqrt{\frac{10^{24}~years}{T^{0\nu -exp}_{1/2}}}.
\end{eqnarray}
$\kappa$ is equal to 1.8 (gluino phenomenological scenario 
\cite{fae98}).

The current experimental upper bounds on the considered
effective LN violating parameters for different
nuclei are shown in Table \ref{table.2}. We see that the 
Heidelberg-Moscow \cite{ge76a} and IGEX \cite{ge76b} experiments 
offer the most restrictive limit for $<m_\nu >$, $\eta_{_N}$
and ${\lambda_{111}'}$, namely  $0.51~eV$, $8.6\times 10^{-8}$
and $1.2\times 10^{-4}$ (assuming 100 GeV masses of SUSY particles),
 respectively. It is interesting to note 
that upper bounds on these parameters from the search for
$0\nu\beta\beta$-decay in $^{134}Xe$ are already stronger or
comparable with those for $^{48}Ca$.

Recently, evidence for the $0\nu\beta\beta$-decay of $^{76}Ge$ 
with half-life $(0.8-18.3)\times 10^{25}$ years (with best value of
$1.5\times 10^{25}$ years) has been reported by the 
Heidelberg group \cite{evid}. 
This work has attracted a lot of attention of both experimentalists and 
theoreticians due to important consequences for the particle physics
and astrophysics. The positive signal implies neutrino is a Majorana
particle and allows to derive patterns of neutrino mass matrix from the 
observed data, if the light neutrino exchange is the dominant mechanism
of this process. By considering matrix elements of Ref. \cite{si99},
which include contributions from higher order terms of the nucleon current, 
one finds for $<m_\nu >$ the value $0.52~eV$. This is partially surprising
as the most favored scenarios of neutrino mixing are preferring 
$<m_\nu >~<~0.01~eV$ \cite{bil99}. It could be that another
mechanism is responsible for this process. However, before making
some strict theoretical conclusions it is important to clarify 
some experimental aspects.  

The main
problem is that the claimed significance of experimental evidence of the 
$0\nu\beta\beta$-decay is not high, around 2 $\sigma$
and that there is not a consensus among various experimental groups
\cite{crit}. It seems that we must wait to see what other experiments find. 
The problem can be solved, if an evidence of the 
$0\nu\beta\beta$-decay will be confirmed  or ruled out by another experiment.
For that purpose we present the calculated half-lifes for nuclei
of experimental interest by assuming the above considered LN violating
mechanisms and that the measurement of ref. \cite{evid} is correct.
They are listed in Table \ref{table.3}. In the near
future the required half-life value can be achieved within the 
NEMO III experiment searching for the $0\nu\beta\beta$-decay in
$^{100}Mo$, which has the chance to reach the level of the
half-life up to $1.\times 10^{25}$ years.

There are some other ambitious project in preparation,
in particular  CAMEO, CUORE, COBRA, ECHO, GENIUS, MAJORANA, 
MOON, XMASS etc \cite{vog02}. The  next generation of the 
$0\nu\beta\beta$-decay detectors will consist of few tons of 
the radioactive $0\nu\beta\beta$-decay 
material.  This is  a great 
improvement as the current experiments use only few tens of
kg's for the source. The above mentioned $0\nu\beta\beta$-decay 
experiments are expected to shed more light on the problem
of Majorana neutrinos within the coming decade. Perhaps, 
there will be a clean evidence of this exotic nuclear 
transition, which will be generally accepted by the 
physical community. If it will be the case, the issue of
nuclear matrix elements will become crucial. We note that
by knowing nuclear matrix elements with high reliability
and the $0\nu\beta\beta$-decay half-lifes for transitions
to $0^+$ excited and ground states, it is possible to 
determine the dominant mechanism \cite{dist}.

In summary, we have shown that $^{134}Xe$ is 
a promising nucleus for examing the $0\nu\beta\beta$-decay. 
The nuclear matrix elements governing this process 
have been found considerably larger in comparison 
with those for $^{136}Xe$. Thus the
sensitivities of both isotopes to the signal of 
effective light Majorana neutrino are close each to other.   
By keeping in mind a comparable abundance of both 
xenon isotopes the results obtained give a strong 
motivation to search simultaneously for the 
$0\nu\beta\beta$-decay of $^{134}Xe$  and $^{136}Xe$. 
Finally, the present status of searches for the
$0\nu\beta\beta$-decay was reviewed. Perspectives
of confirming or ruling out the evidence 
for the $0\nu\beta\beta$-decay of $^{76}Ge$ were
discussed.

This work was supported in part by the Deutsche 
Forschungsgemeinschaft (436 SLK 17/298) and by the 
Grant agency of the Czech Republic under contract 
No. 202/02/0157.


\begin{table}[t]
\caption{Nuclear matrix elements of light and heavy Majorana neutrino 
exchange and trilinear R-parity violating modes
for the $0\nu\beta\beta$-decay in $^{134}Xe$ and $^{136}Xe$.
}
\label{table.1}
\begin{tabular}{lccccccccc}
 & \multicolumn{4}{c}{Light neutrino exch. mech.} & & 
\multicolumn{4}{c}{Heavy neutrino exch. mech.}\\ \cline{2-5} \cline{7-10}
 Nucl. &
$M^{<m_\nu >}_F$ & $M^{<m_\nu >}_{GT}$ & $M^{<m_\nu >}_T$ & 
${\cal M}_{<m_\nu >}$ &  &
$M^{\eta_{_N}}_F$ & $M^{\eta_{_N}}_{GT}$ & $M^{\eta_{_N}}_T$ & 
${\cal M}_{\eta_{_N}}$ \\ \hline
$^{136}Xe$ & 
 -0.981   & 1.286    & -0.252    &  1.66   &  &
 -35.9    &  27.7    & -27.7     &  23.0      \\
$^{136}Xe$ & 
 -0.504   & 0.496    & -0.161    &  0.66   &  &
 -21.7    &  16.8    & -16.6     &  14.1      \\\hline
 & & &
\multicolumn{4}{c}{$R_p \hspace{-1em}/\;\:$  SUSY mech.}\\ \cline{2-10} 
   &
 ${M}_{GT}^{1\pi}$ & ${M}_{T}^{1\pi}$ & ${\cal M}_{}^{1\pi}$ & &
 ${M}_{GT}^{2\pi}$ & ${M}_{T}^{2\pi}$ & ${\cal M}_{}^{2\pi}$ & &
 ${\cal M}_{{\lambda'}_{111}}$ \\ \hline
 $^{134}Xe$ & 
   1.078    &  -1.417 & 30.0 & &  -1.237 & -0.899 & -644. & &    -614. \\
 $^{136}Xe$ & 
   0.606 &  -0.840 & 20.7 & &  -0.742 & -0.543 & -387. & &    -367. \\
\end{tabular}
\end{table}

\begin{table}[t]
\caption{The present state of the Majorana neutrino mass (light and heavy),
and $R_p \hspace{-1em}/\;\:$ SUSY searches
in $0\nu\beta\beta$-decay experiments. 
$T^{exp-0\nu}_{1/2}$ is the best presently available lower limit on
the half-life of the $0\nu\beta\beta$-decay  to the ground state
for a given isotope.
$\zeta_{<m_\nu >} (Y)$, $\zeta_{\eta_N} (Y)$ and 
$\zeta_{{\acute{\lambda}}_{111}} (Y)$ 
denote according to Eq. (\protect\ref{eq:6}) the sensitivity of
a given nucleus $Y$ to the light neutrino mass, heavy neutrino mass,
and $R_p \hspace{-1em}/\;\:$ SUSY signals, respectively. 
The corresponding upper limits on LN 
non-conserving parameters ${<m_\nu >}$, $\eta_{N}$ 
and $\lambda_{111}'$ (assuming mass of SUSY particles of 100 GeV) 
are presented. 
}
\label{table.2}
\begin{tabular}{lrrrrrrrrrrr}
 Nucleus    & $^{48}Ca$  & $^{76}Ge$  & $^{82}Se$  & $^{96}Zr$ & $^{100}Mo$ &
 $^{116}Cd$ & $^{128}Te$ & $^{130}Te$ & $^{134}Xe$ & $^{136}Xe$ & $^{150}Nd$ 
\\ \hline
$T^{exp-0\nu}_{1/2}$ & 
$ 9.5\times $ & $ 1.6\times $ & $ 1.4\times $ & $ 1.0\times $ & 
$ 5.5\times $ & $ 7.0\times $ & $ 8.6\times $ & $ 1.4\times $ &
$ 5.8\times $ & $ 7.0\times $ & $ 1.7\times $   \\
$[years]$  & $10^{21}$ & $10^{25}$ & $10^{22}$ & $10^{21}$ & 
$10^{22}$ & $10^{22}$ & $10^{22}$ & $10^{23}$ & $10^{22}$ & 
$10^{23}$ & $10^{21}$ \\
C.L. [\%] & 80 & 90 & 90 & 90 & 90 & 90 & 90 & 90 & 90 & 90 & 95 \\
Ref. & 
\cite{ca48}  & \cite{ge76a,ge76b} & \cite{se82}  & \cite{zr96}   & 
\cite{mo100} & \cite{cd116}       & \cite{te128} & \cite{te128}  & 
\cite{xe134} & \cite{xe136}       & \cite{nd150}   \\
 & & & & & & & & & & &\\
$\zeta_{<m_\nu >} $ &
 2.32 & 2.49 &  4.95 & 4.04 &  7.69 &  5.11 &
 1.02 &  4.24 & 0.80 & 1.60 &  17.3 \\
$\zeta_{\eta_{_N}} $  &
 0.48 & 2.90 &  5.64 & 3.98 &  7.10 &  5.36 &
 1.25 &  5.45 &  1.10 & 3.43 &  18.5 \\
 $\zeta_{\lambda_{111}'} $  &
 4.18 & 5.57 &  10.9 & 11.6 & 17.9 &  10.9 & 
 3.25 & 14.7 & 2.94 & 8.92 & 54.7 \\
 & & & & & & & & &  & &\\
 $<m_\nu >~[eV]$  & 
 22. & 0.51 & 8.7 & 40. & 2.8 & 3.8 & 17. & 3.2 & 27. &
 3.8 & 7.2  \\
 $\eta_{_N}~[10^{-7}]$ & 
 $2~ 10^2$ & $0.86$ & $15.$ & $79.$ &
$6.0$ & $7.0$ & $27.$ & 
$4.9$ &  $38.$ & $3.5$ & $13.$ \\   
$\lambda_{111}'~[10^{-4}]$  &
 8.9 & 1.2 & 5.0 & 9.4 & 2.8 & 3.4 & 
5.8 & 2.4 & 6.8 & 2.1 & 3.8 \\
\end{tabular}
\end{table}

\begin{table}[t]
\caption{Theoretical half-lifes of the $0\nu\beta\beta$-decay 
to the ground state for nuclei of experimental interest 
by assuming, that the evidence of the
$0\nu\beta\beta$-decay of $^{76}Ge$ with 
$T^{0\nu-theor.}_{1/2}(0^+_{g.s.}\rightarrow 0^+_{g.s.})~ =~
 1.5 \times 10^{25}~ years$  \protect\cite{evid} is correct.
l.n.e and h.n.e. stand for light and heavy Majorana neutrino exchange,
respectively.
}
\label{table.3}
\begin{tabular}{lrrrrrrrrrr}
 & \multicolumn{10}{c}{ 
$T^{0\nu-theor.}_{1/2}(0^+_{g.s.}\rightarrow 0^+_{g.s.})$ [years]} 
\\\cline{2-11}
    & $^{48}Ca$  & $^{82}Se$  & $^{96}Zr$ & $^{100}Mo$ &
 $^{116}Cd$ & $^{128}Te$ & $^{130}Te$ & $^{134}Xe$ & $^{136}Xe$ & $^{150}Nd$ 
\\ 
mech.  & 
$\times 10^{25}$ & $\times 10^{24}$ & $\times 10^{24}$ & $\times 10^{24}$ & 
$\times 10^{24}$ & $\times 10^{25}$ & $\times 10^{24}$ & $\times 10^{26}$ &
$\times 10^{25}$ & $\times 10^{23}$ \\ \hline
l.n.e. & 
1.7 & 3.8 & 5.7 & 1.6 & 3.6 & 9.0 & 5.2 & 1.5 & 3.6 & 3.1 \\
h.n.e. &
5.5 & 4.0 & 8.0 & 2.5 & 4.4 & 8.1 & 4.3 & 1.0 & 1.1 & 3.7 \\
$R_p \hspace{-1em}/\;\:$ SUSY &
2.7 & 3.9 & 3.5 & 1.4 & 3.9 & 4.4 & 2.1 & 54. & 59. & 1.6 \\
\end{tabular}
\end{table}

\end{document}